\documentclass[pre, reprint,floats,amsmath,amssymb,superscriptaddress,floatfixl,aps,longbibliography]{revtex4-2}
\usepackage[utf8]{inputenc}
\usepackage{amsmath,amssymb}
\usepackage[colorlinks=true,citecolor=red,filecolor=green,linkcolor=blue,pdfnewwindow=true]{hyperref}

\usepackage{graphicx} 
\usepackage{physics}
\usepackage{xcolor}

\begin{document}
\title{Power spectral density of trajectories of active Ornstein--Uhlenbeck particles}

\author{Yeongjin Kim}
\affiliation{Department of Physics, Pohang University of Science and Technology (POSTECH), Pohang 37673, Republic of Korea}

\author{Gleb Oshanin}
\email{gleb.oshanin@gmail.com}
\affiliation{Sorbonne Universit\'e, CNRS, Laboratoire de Physique Th\'eorique de la Mati\'ere
Condens\'ee (UMR CNRS 7600), 4 Place Jussieu, 75252 Paris Cedex 05, France}
\affiliation{Asia Pacific Center for Theoretical Physics (APCTP), Pohang 37673, Republic of Korea}

\author{Jae-Hyung Jeon}
\email{jeonjh@gmail.com}
\affiliation{Department of Physics, Pohang University of Science and Technology (POSTECH), Pohang 37673, Republic of Korea}
\affiliation{Asia Pacific Center for Theoretical Physics (APCTP), Pohang 37673, Republic of Korea}


\begin{abstract}
The power spectral density (PSD) is a central frequency-domain descriptor of stochastic processes. While PSDs have been studied for Brownian motion and a few anomalous diffusion processes, the spectral densities of active nonequilibrium processes remain almost unexplored. Here, we present an exact theory for the PSDs of active diffusion using the model of active Ornstein--Uhlenbeck particles (AOUPs).  We investigate the spectral densities of AOUPs in free space and under harmonic confinement. 
In free space, active motion does not alter the Brownian $f^{-2}$ spectrum, but only modifies its amplitude and introduces a crossover at the persistence frequency. 
Under confinement, the spectrum exhibits a rich variety of features depending on the persistence, trap relaxation, and activity strength, including two characteristic signatures that are absent in both thermal systems and free AOUPs.
These are a two-plateau structure from a double-trapping mechanism due to two noise sources, and the new $f^{-4}$ spectral scaling associated with transient ballistic motion. 
We also investigate the finite time effects through the finite-time PSD, and find that the low-frequency plateau and high frequency oscillation exhibit distinct dependences on the observation time $T$ in free and confined systems. 
Finally, we discuss our results in connection with previously reported experimental studies of active systems. Our results provide an analytically tractable framework for interpreting such systems. 
\end{abstract}

\maketitle

\section{Introduction}

The power spectral density (PSD) is a fundamental characteristic of time-series data and stochastic processes, providing essential insight into their frequency-domain behavior \cite{norton}. Its central role originates from the classical theory of Brownian motion and the Lorentzian spectrum of the Ornstein–Uhlenbeck process, which directly encodes equilibrium fluctuation–dissipation relations \cite{uhlenbeck1930theory, kubo1966fluctuation}. In equilibrium systems, the PSD serves as a distinctive spectral signature that enables the extraction of key system parameters, such as the stiffness of a harmonic trap in optical tweezer experiments \cite{berg2004power}.

While Brownian motion remains the paradigmatic model with well-understood spectral properties (see, e.g., \cite{krapf2018power}), many natural systems exhibit anomalous dynamics, characterized by pronounced deviations from Brownian behavior arising from correlated noise, environmental disorder, or underlying velocity fields \cite{ralf1,ralf2}. Consequently, recent work has focused on fundamental aspects of the power spectral densities of anomalous diffusion processes \cite{mandel,lowen,watkins,enzo,eli2,eli}, including the full probability distributions of single-trajectory PSDs and their frequency–frequency correlations \cite{luca1,luca2}, as well as detailed analyses of several representative models. These include, to name a few, fractional Brownian motion \cite{krapf2019spectral,sposini2}, scaled Brownian motion \cite{sposini2019single}, diffusing diffusivity processes \cite{sposini}, systems in the universality class of the Kardar–Parisi–Zhang equation \cite{takeuchi}, and trajectories of membrane proteins \cite{eli3}.

Over recent decades, the dynamics of intrinsically nonequilibrium systems -- most prominently those arising in active matter -- have become a central topic of research. Compared to equilibrium systems, such systems often display markedly richer and more complex dynamical behavior. Active particles, such as, e.g., Janus colloids, motor proteins, and biological swimmers, continuously consume energy from their environment to generate self-propulsion, leading to non-thermal fluctuations with finite correlation times \cite{bechinger2016active,song2018neuronal,song2023machine}. Despite growing experimental interest, the spectral properties of active trajectories remain largely unexplored, with only recent work \cite{solon} addressing the power spectral density of active Brownian particle motion.

In this paper, we extend the analysis of Ref.~\cite{solon} by focusing on the active Ornstein--Uhlenbeck particle (AOUP) model--a minimal yet analytically tractable framework that has attracted considerable attention in recent years~\cite{szamel2014self,maggi2015multidimensional,fodor2016far,wittmann2017_jstat113207,bonilla2019_aoup,dabelow2021irreversibility,martin2021_pre032607}. The AOUP model incorporates exponentially correlated active forces with a persistence time 
$\tau_A$, and allows analytical calculation of the steady-state distributions, effective temperatures, and the statistics of individual trajectories. Despite the extensive body of work on AOUP dynamics, the power spectral density -- particularly the finite-time PSD relevant to experiments--has not yet been fully characterized. A comprehensive understanding of how activity strength, persistence, and confinement shape the PSD is therefore essential, both for interpreting experimental spectra and for establishing PSD analysis as a quantitative probe of active fluctuations.

This paper is  organized as follows: 
In Sec.~\ref{sec2}, we introduce the definition of PSDs and the basic results relevant to this work. In Sec.~\ref{sec3}, we present the AOUP model in free and harmonically confined space. Their important dynamic properties, such as the position autocovariance and the mean squared displacement, are explained and used in the subsequent two main sections. In Secs.~\ref{sec4}~and~\ref{sec5}, we present our anaytic and numerical studies of the power spectral densities of free and harmonically confined AOUPs. Finally, in Sec.~\ref{sec6}, we summarize the main findings and discuss their applications to some relevant experimental systems in the literature.

\section{Power spectral density\label{sec2}}

In the classical theory of stochastic processes, the power spectral density (PSD) of a stationary random process $X(t)$ is defined as the Fourier transform of its autocovariance function \cite{norton}. For a wide-sense stationary process with finite second moments, the Wiener--Khinchin theorem states that the PSD $\mu(f)$, where $f$ is the frequency, is given by the Fourier transform of the time-lag autocovariance
\begin{equation}
	C_X(\tau)=\langle X(t)X(t+\tau)\rangle-\langle X(t)\rangle^{2},
\end{equation}
where $\langle\cdot\rangle$ denotes a full ensemble average. Equivalently, the PSD can be expressed as the infinite-time limit of the squared modulus of the Fourier transform of $X(t)$,
\begin{equation}
	\mu(f,\infty)=\lim_{T\to\infty}\frac{1}{T}\,
	\Biggl\langle
	\Biggl|
	\int_{0}^{T} X(t)\,e^{-i f t}\,\mathrm{d}t
	\Biggr|^{2}
	\Biggr\rangle \,.
\end{equation}
For ergodic stationary processes, this ensemble average may be replaced by a time average over a single infinitely long trajectory.

In the standard theoretical analysis, the limit 
$T \to \infty$ plays a central role: it guarantees stationarity of the spectral estimate, arbitrarily fine frequency resolution, and the equivalence between ensemble-based and time-averaged definitions of the power spectral density. In practice, however, neither experiments nor numerical simulations have access to infinite observation times and, in many cases, only a limited number of realizations are available. Moreover, only a restricted class of stochastic processes encountered in nature is strictly stationary.

For nonstationary dynamics, the conventional definition of the PSD as the Fourier transform of the autocovariance function either does not exist or fails to yield meaningful results in the infinite-time limit \cite{eli1,eli2}. A prominent example is fractional Brownian motion with Hurst exponent 
$H>1/2$ \cite{mandel}, for which the process is superdiffusive and nonstationary. In this case, the standard Wiener–Khinchin formalism breaks down, as the autocorrelation structure leads to divergences in the spectral integral \cite{krapf2019spectral}. 

In this case, one either resorts to some alternative definition of the PSD (see, e.g., \cite{SquarciniMarinariOshanin2020}) or focuses on  the finite-time PSD of the form
\begin{equation}
	\label{main}
	\mu(f,T)=\frac{1}{T}\,
	\Biggl\langle
	\Biggl|
	\int_{0}^{T} X(t)\,e^{-i f t}\,\mathrm{d}t
	\Biggr|^{2}
	\Biggr\rangle  \,,
\end{equation}
in which the constraint $T \to \infty$ is relaxed. This finite-time definition has proven to be particularly useful for the analysis of stochastic trajectories obtained in experiments and numerical simulations, where observation times are necessarily limited. It yields well-defined and physically meaningful spectral estimates, allowing one to characterize both the frequency dependence of fluctuations and the non-stationary features of the dynamics. In particular, the explicit dependence of the PSD on the observation time $T$ encodes aging effects and reveals large-$T$ divergences that cannot be captured within the conventional infinite-time framework, thereby providing direct access to dynamical signatures of non-equilibrium and anomalous transport processes \cite{krapf2019spectral,sposini2}.
 
Lastly, we note that for one-dimensional real-valued processes the expression \eqref{main} can be simply rewritten as
\begin{equation}
    \mu(f,T) 
    = \frac{1}{T}\int_{0}^{T}\!dt_1 \int_{0}^{T}\!dt_2\,
    \cos\!\big(f(t_1-t_2)\big)\, C_X(t_1,t_2).
    \label{eq:finitePSDReal}
\end{equation}
where $C_X(t_1,t_2)\equiv \langle X(t_1)X(t_2)\rangle$ is the two-time correlation (covariance) function  of the process $X(t)$. 
Equation~\eqref{eq:finitePSDReal} shows that 
the finite-time PSD is explicitly expressed by the cosine transform of the temporal correlations of $X(t)$.

Having introduced the finite-time power spectral density in Eq.~\eqref{eq:finitePSDReal}, we now turn to a systematic analysis of its dependence on both the frequency and the observation time for the active Ornstein–Uhlenbeck particle (AOUP) model, which will be defined in detail below. This model extends classical thermal Brownian motion by incorporating an exponentially correlated active force, thereby providing a minimal description of self-propelled dynamics. We consider two representative situations: free motion, which leads to intrinsically non-stationary dynamics, and motion in a harmonic potential, for which a stationary regime is attained at long times. These two settings allow us to illustrate how activity and confinement shape the spectral properties of stochastic trajectories in both non-equilibrium and steady-state conditions.

\section{Active Ornstein-Uhlenbeck particles\label{sec3}}

To analyze how activity reshapes the power PSD of particle trajectories, we here employ the so-called active Ornstein--Uhlenbeck particle (AOUP) model among several popular models of self-propelled particles. This model provides a minimal yet analytically tractable description of self-propelled motion for biological or artificial active entities~\cite{bechinger2016active}. Its linear Gaussian structure allows us to evaluate autocorrelation functions, making it suitable for spectral analysis. 
The PSD for other active diffusion models, such as active Brownian particle \cite{fodor2018statistical, sevilla2015smoluchowski,kurzthaler2018probing} and run-and-tumble \cite{martens2012probability} motion, will be discussed in the discussion section.

In free space (in the absence of confinement or spatially inhomogeneous effects), the AOUP is described by the following coupled Langevin equations~\cite{semeraro2021work,dabelow2021irreversibility}:
\begin{align}
    \zeta \frac{dx}{dt} & = \xi (t) + F (t), 
    \label{FreeAOUPLangevin}
    \\
    \frac{d F}{dt} & = - \frac{1}{\tau_A} F + \sqrt{\frac{2 \zeta ^2 v_p ^2}{\tau _A}} \eta (t), 
    \label{AOUPactiveForceLangevin}
\end{align}
where $\zeta$ is the friction coefficient of the medium and $\xi(t)$ is thermal Gaussian noise satisfying the fluctuation--dissipation theorem: $\langle \xi(t) \rangle = 0$ and $\langle \xi(t) \xi(t') \rangle = 2 \zeta k_B \mathcal{T} \delta (t-t')$ where $k_B$ is the Boltzmann constant and $\mathcal{T}$ is the absolute temperature.
The self-propulsion motion is driven by $F(t)$, an Ornstein--Uhlenbeck (OU) force with a persistence time $\tau_A$. The noise $\eta(t)$ is Gaussian white noise with unit variance.  Assuming that the active OU process has been running from $t=-\infty$ such that the initial condition is neglected, the active force $F(t)$ is stationary and exponentially correlated,
\begin{equation}
\begin{aligned}
    \langle F(t) F(t') \rangle  = \zeta ^2 v_p ^2 e^{- \qty| t-t'|\slash \tau_A}.
\end{aligned}
\end{equation}
Here, $v_p$ defines the propulsion speed of the active motion, while $\tau_A$ is the active persistence time. Although $F(t)$ is stationary, the particle position $x(t)$ is non-stationary.

\begin{figure}
    \centering
    \includegraphics[width=0.45\textwidth]{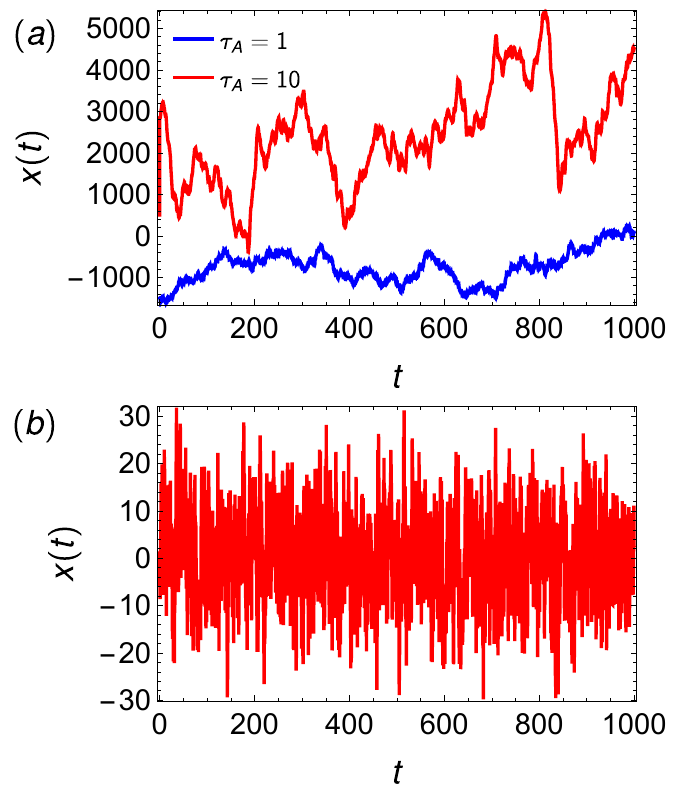}
\caption{\label{Fig1}
Sample trajectory of AOUPs. (a) Two simulated trajectories for free AOUPs with $\tau_A = 1$ and $10$ at $v_p =10$.  (b) A simulated trajectory for an AOUP in a harmonic potential with stiffness $k=1$ for given $\tau_A = 1$ and $v_p =10$.}
\end{figure}

Solving Eq.~\eqref{FreeAOUPLangevin} and \eqref{AOUPactiveForceLangevin} yields the two-time position autocovariance function of the AOUP:
\begin{align}
\begin{aligned}
C_{x}(t,t') &= \frac{2 k_B \mathcal{T}}{\zeta} \min \qty(t, t') + 2 v_p ^2 \tau_A  \min \qty(t, t')  \\
 &\hspace{-2em} - v_p ^2 \tau_A ^2 \left(1 + e^{-\tfrac{ \qty|t-t'|}{\tau_A}}  - e^{-\tfrac{\min \qty(t, t')}{\tau_A}} - e^{- \tfrac{\max \qty(t, t')}{\tau_A} }\right).
 \end{aligned}
\label{CxxFreeAOUP}
\end{align}
The first term in the R.H.S. is the well-known autocorrelation form of a thermal Wiener process (Brownian motion). The other two terms are attributed to the active dynamics from $F(t)$. 
In particular, the last term captures the correlation effect of the exponentially decaying active OU force. 
Setting $t=t'$ gives the mean-squared displacement  $\langle \Delta x^2(t)\rangle\equiv\langle [x(t)-x(0)]^2\rangle$ for AOUPs~\cite{howse2007self}, 
\begin{equation}
\begin{aligned}
    \langle \Delta x^2 (t) \rangle & = \frac{2k_B \mathcal{T}}{\zeta} t + 2v_p ^2 \tau_A t - 2v_p ^2 \tau_A ^2 \qty(1-e^{-t\slash \tau_A}), \\
    & = \begin{cases}
       2Dt  & t \ll  \frac{2D}{v_p ^2}  \\
       v_p^2 t^2 & \frac{2D}{v_p ^2}  \ll t \ll \tau_A \\
       2(D + v_p^2\tau_A )t & t\gg \tau_A 
    \end{cases}.
\end{aligned}
\label{eq:aoupmsd}
\end{equation}
This expression reveals three distinct dynamical regimes. For short times of $t\ll 2D/v_p^2$, the thermal motion dominates over the directional motion, in which the MSD is governed by Einstein's diffusion law with the diffusivity $D=k_B\mathcal{T}/\zeta$.  
For intermediate times of $2D/v_p^2\ll t\ll \tau_A$, the persistent propulsion produces ballistic motion $\Delta x(t)\simeq v_p t$. For long times of $t\gg \tau_A$, the active force decorrelates and the AOUP becomes Fickian again with the effective diffusivity
\begin{equation}
     D_{\mathrm{eff}} = D + D_A,
     \label{Deff}
\end{equation}
which is the sum of the thermal diffusivity and the activity-induced diffusivity 
\begin{equation}
     D_A \equiv v_p ^2 \tau_A. 
\end{equation}
The AOUP dynamics in this regime is often referred to as active Fickian in the sense that the diffusivity $D_\mathrm{eff}$ does not satisfy the Einstein relation~\cite{golestanian2009anomalous,fodor2016far}. 
The relative strength of activity is conveniently characterized by the dimensionless ratio
\begin{equation}
    \mathrm{Pe}\equiv \frac{D_A}{D}.
\end{equation}
For illustration, in Fig.~\ref{Fig1}(a), we show the simulated trajectories of AOUPs with two distinct values of $\tau_A$ for a given $v_p$ and $D$. 

We now consider the AOUP in a harmonic potential $U=\frac{1}{2}kx^2$, so that
\begin{align}
    \zeta \frac{dx}{dt} & = -k x + \xi(t) + F(t), 
    \label{eq:AOUP_harmonic_x}
\end{align}
where $k$ is the potential stiffness, and $\xi$ and $F$ are, respectively, the thermal and active forces explained above. A sample trajectory of the harmonically confined AOUP is depicted in Fig.~\ref{Fig1}(b). This model has been studied in the context of colloids in active baths under confinement~\cite{maggi2014generalized, maggi2017memory}, trapped-and-hopping diffusion of active tracers in a polymer network~\cite{kim2024active,woillez2020active}, or tagged monomer dynamics of a single polymer in active baths~\cite{goswami2022reconfiguration, han2023nonequilibrium}, and a statistical model for the work fluctuations or thermodynamic uncertainty relations of active particles~\cite{semeraro2023work, han2025thermodynamic}.

In contrast to the free case, the confined system reaches a stationary steady state. The deterministic part of the equation, $\zeta \langle \dot{x}\rangle  = -k \langle x\rangle$, yields an exponential relaxation over the timescale
\begin{align}
\tau_R = {\zeta}\slash {k}.
\end{align}
The interplay between the active persistence time $\tau_A$ and the trap relaxation time $\tau_R$ is the central control parameter of the problem.

In the stationary state, the position autocovariance depends only on the time difference and reads~\cite{goswami2022reconfiguration} 
\begin{align}
\begin{aligned}
C_{x}(t-t') &=
 \frac{k_B \mathcal{T}}{k} e^{ - \left | t-t' \right |\slash \tau_R}
\\
&+ \frac{D_A \tau_R ^2 }{\tau_A ^2 - \tau_R ^2}
\left(
\tau_A e^{-\frac{\qty|t - t'| }{\tau_A} }  - \tau_R e^{-\frac{\qty|t - t'| }{\tau_R} } 
\right).
\end{aligned}
\label{activeAutocorrelationFtn}
\end{align}
This expression has a clear structure. The first term is the equilibrium Ornstein--Uhlenbeck result. The second term reflects active persistence and contains two exponential contributions with timescales $\tau_A$ and $\tau_R$. 

The stationary-state variance follows from $C_{x}(0)$, 
\begin{equation}
    \langle x^2 \rangle
    = \frac{k_B \mathcal{T}}{k} + \frac{D_A \tau_R ^2 }{\tau_A + \tau_R}.
    \label{eq:AOUP_variance}
\end{equation}
Since the steady-state distribution of $x$ remains Gaussian under harmonic confinement, we can define the effective temperature of the system via $\langle x^2 \rangle ={k_B \mathcal{T}_{\mathrm{eff}}}\slash {k}$, which yields
\begin{equation}
    \mathcal{T}_{\mathrm{eff}} \equiv \mathcal{T} + \frac{k D_A \tau_R^2}{k_B(\tau_A + \tau_R )}.
    \label{Eq:effectivetemp}
\end{equation}
This effective temperature characterizes the enhanced fluctuations due to activity, although the system remains out of equilibrium. 
The stationary MSD is~\cite{kim2024active}
\begin{equation}
    \begin{aligned}
        \langle \Delta x ^2 (t) \rangle & =
         2 D \tau_R  \qty(1-e^{-t\slash \tau_R}) \\
        & + 
         2\frac{D_A\tau_R ^2 }{\tau_A ^2 - \tau_R ^2}
    \left(
     \tau_A \qty(1-e^{-\frac{t}{\tau_A} })- \tau_R \qty(1-e^{-\frac{t}{\tau_R}} )
    \right).
    \end{aligned}
    \label{Eq:HarmonicMSD}
\end{equation}
Consistent with the definition of the effective temperature [Eq.~\eqref{Eq:effectivetemp}], the MSD approaches to the saturation $\langle \Delta x^2 \rangle
= {2k_B \mathcal{T}_{\mathrm{eff}}}\slash{k}$ as time goes to infinity. The MSD displays rich dynamic patterns depending on the values of the two characteristic timescales $\tau_A$ and $\tau_R$ and $\mathrm{Pe}=D_A/D$. Further discussion of the MSD is given in Sec.~\ref{secPSDandMSD}.
In the following sections, we demonstrate how the hierarchy between $\tau_A $, $\tau_R$, and the activity strength governs the structure of the PSD for both infinite and finite observation time.

\section{Power spectral densities for free AOUPs\label{sec4}}
In this section, we investigate the PSDs for the AOUP model governed by the Langevin equations~\eqref{FreeAOUPLangevin} and \eqref{AOUPactiveForceLangevin}. 
We analytically derive the spectral densities using the position autocovariance, supplement the results with numerical simulations, and explore how the scalings and amplitudes of the PSDs depend on the activity-induced correlated diffusion and on the observation time (or trajectory length) $T$.

\begin{figure}
\includegraphics[width=0.45\textwidth]{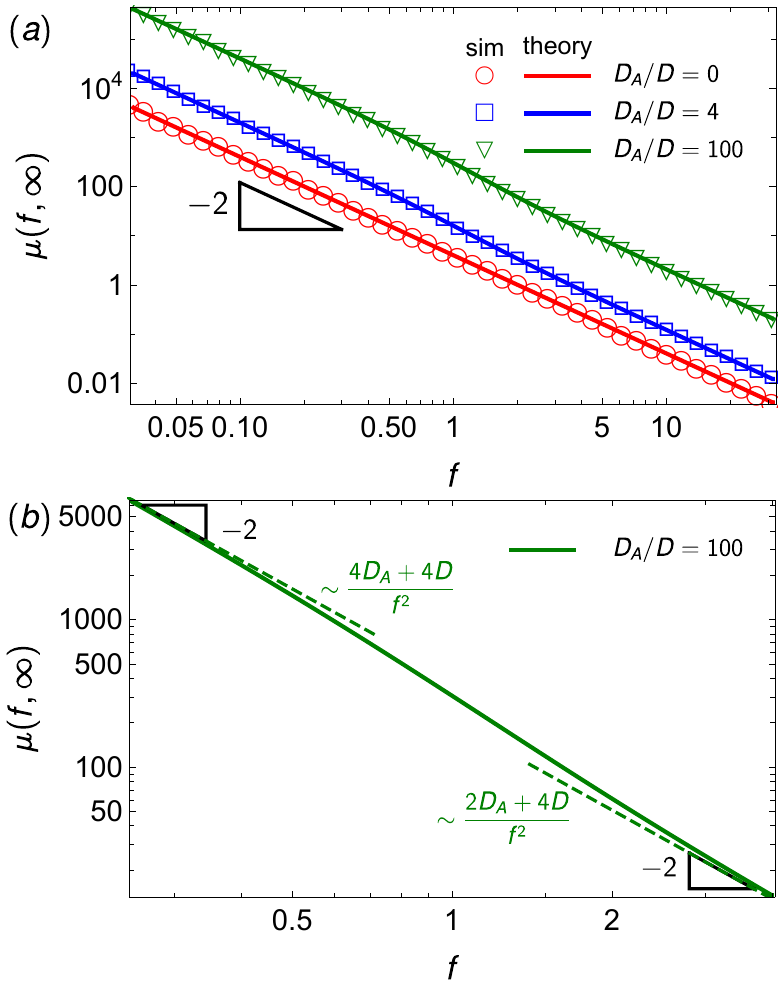}
\caption{\label{Fig2} 
{Infinite-$T$} PSDs of the free-space AOUP model. (a) Theoretical PSD~\eqref{activeFreePSD} for increasing the activity strength Pe, compared with the numerical results (symbols). The numerical data is obtained via Eq.~\eqref{main} with Eqs.~\eqref{FreeAOUPLangevin}~and~\eqref{AOUPactiveForceLangevin} simulated trajectories of length $T=1000$ for given parameter values. (b) The zoom-in PSD~\eqref{activeFreePSD}, together with the two limiting asymptotics (the dashed lines).
}
\end{figure}

\subsection{{ Infinite-$T$} PSD}
As the free space motion is nonstationary, we evaluate the PSD using the cosine transformation of the two-point position correlation function through Eq.~\eqref{main}. Using the corresponding position autocovariance derived in Eq.~\eqref{CxxFreeAOUP}, we obtain 
\begin{equation}
\begin{aligned}
    \mu(f,\infty) 
    & = \frac{4D + 4 D_A }{f^2} -  \frac{2D_A}{f^2 + \tfrac{1}{\tau_A ^2}}.
    \label{activeFreePSD}
\end{aligned}
\end{equation}
When the activity is removed ($D_A\to 0$), the PSD recovers the well-known PSD for Brownian motion, $\mu(f,\infty) = \frac{4D}{f^2}$. 
Activity-induced motion yields an additional Lorentzian correction that interpolates the prefactor beyond this free Brownian PSD. 
Specifically, the spectrum crosses over from $\mu(f,\infty)\simeq 4D_{\mathrm{eff}}/f^{2}$ for $f\ll\tau_A^{-1}$ to $\mu(f,\infty)\simeq (2D+2D_{\mathrm{eff}})/f^{2}$ for $f\gg\tau_A^{-1}$.
To confirm our analytic result, we numerically simulate the AOUP model and compute PSDs from the simulated time series  (Appendix~\ref{label:sim} for simulation details).

This is shown in Fig.~\ref{Fig2}(a), where the numerical simulation (symbols) is in excellent agreement with the theoretical result, Eq.~\eqref{activeFreePSD}, in the wide range of $f$ at varying activity strengths.

A key feature observed in the plot is that the free AOUP PSD retains the diffusive scaling $\mu \propto f^{-2}$ over the entire frequency range. The activity does not introduce an additional high-frequency power-law scaling distinct from $f^{-2}$. Instead, activity modifies the PSD amplitudes with a crossover around $f\sim \tau_A^{-1}$.

In the low-frequency regime, $f\ll\tau_A^{-1}$, the active force is effectively decorrelated on the corresponding long timescales. Consequently, the active diffusion converges to the Fickian diffusion with the aforementioned effective diffusivity in Eq.~\eqref{Deff}, yielding
\begin{equation}
    \mu(f,\infty) \approx 
    \frac{4D_{\mathrm{eff}}}{f^2}
    . 
    \label{PSDfreeAOUPsmallf}
\end{equation}
See the zoom-in plot of the PSD \eqref{activeFreePSD} in Fig.~\ref{Fig2}(b).  The same dynamic behavior can be observed in the MSD at this timescale, see Eq.~\eqref{eq:aoupmsd}. This result is intuitively consistent with the thermal Brownian result $\mu= 4Df^{-2}$: at long times, activity simply enhances diffusion, so the diffusivity in the power spectrum is replaced by the effective diffusivity. 

More intriguingly, activity leaves a clear signature in the high-frequency regime $f\gg \tau_A^{-1}$ as in Fig.~\ref{Fig2}(b), where the PSD again scales as $f^{-2}$ but with a different amplitude,
\begin{equation}
\mu(f,\infty) \approx \frac{2D + 2D_{\mathrm{eff}}}{f^2}.
\label{PSDfreeAOUPlargef}
\end{equation}

This is indeed a nontrivial result, which is not obvious at all from real-time observables such as the MSD. At sufficiently short lag times, the MSD is dominated by the thermal fluctuation $\langle \Delta x^2(t)\rangle \simeq 2Dt$, whereas the persistent active contribution $\sim v_p^{\,2}t^{2}$ is subleading and cannot be detected for $t\ll 2 \left(\frac{D}{D_A}\right)\tau_A$. Although the active entities behave as thermal Brownian motion at this timescale, its counterpart PSD does not converge to $4D f^{-2}$. Surprisingly, the PSD retains the active contribution throughout the whole frequency range.
Therefore, the PSD provides a sharper probe of active motion than the usual MSD.

\begin{figure}
    \includegraphics[width=0.45\textwidth]{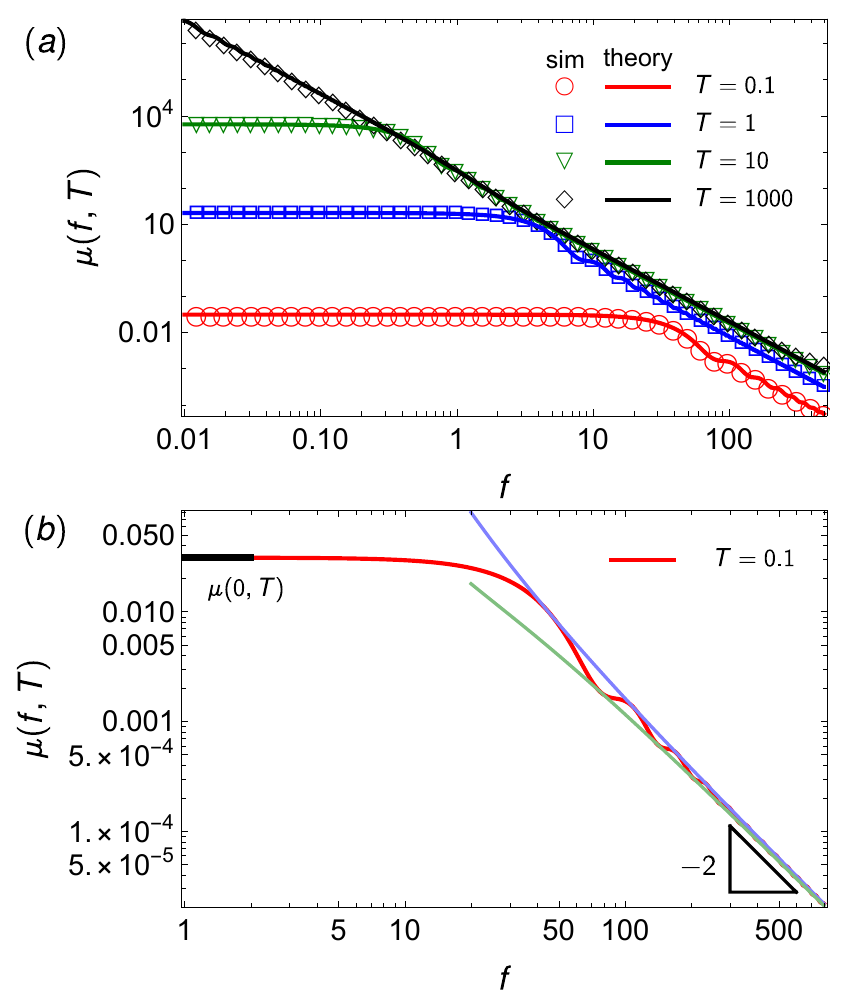}
    \caption{\label{Fig3}
    (a) {Finite-$T$} PSD $\mu(f,T)$ of a free AOUP with $v_p=10$, $\tau_A=1$, and $\zeta=1$ for a short window time $T=0.1$. Symbols are simulations and solid lines are the analytical result, Eq.~\eqref{Eqn:SingleFreeAOUPPeriodogram}.
    (b) Zoom-in showing the oscillations induced by the finite observation window of $T=0.1$, together with guide lines indicating the oscillation amplitude. 
    }
\end{figure}

\subsection{Finite-$T$ PSD}
We next investigate the finite-$T$ PSD, $\mu (f,T)$, evaluated using finite trajectories of length $T$. The exact closed-form expression for $\mu (f,T)$ can be obtained by performing the integral~\eqref{eq:finitePSDReal} with the two-time position autocovariance $C_{x}(t,t')$ shown in Eq.~\eqref{CxxFreeAOUP}.  The exact analytic expression for $\mu (f,T)$ is provided in Eq.~\eqref{Eqn:SingleFreeAOUPPeriodogram} (Appendix~\ref{sec:freeperiodogram}). 

In Fig.~\ref{Fig3}(a), we plot our theoretical PSD curves (lines), Eq.~\eqref{Eqn:SingleFreeAOUPPeriodogram}, together with the numerical results (symbols) for various trajectory length $T$. Compared to the infinite-time PSD above, the finite-time PSD exhibits two characteristic finite-window features: The PSD $\mu (f,T)$ has a low-frequency plateau as $f\to 0$ and shows a weak oscillation in the high-frequency regime while overall keeping the $f^{-2}$ scaling. These two features are shown in both simulated PSDs and our theoretical curves with excellent agreement. 

To analyze these behaviors in detail, in Fig.~\ref{Fig3}(b) we replot our theoretical PSD curve $\mu(f,T=0.1)$ within an appropriate frequency range. Our theoretical PSD shows that the zero-frequency plateau approaches the asymptotic
\begin{equation}
\begin{aligned}
    & \mu (f\to0,T)  =
    \frac{2 D T^2}{3 }
    \\ &  \quad \quad
    +\frac{D_A \left(2
   T^3-3 T^2 \tau_A-6 \tau_A^2 e^{-\frac{T}{\tau_A}} (T+\tau_A)+6 \tau_A^3\right)}{3 T}.
\end{aligned}
\label{eq:S0freeAOUPPSD}
\end{equation}
(In the plot, $\mu (f=0,T)$ is annotated with the thick solid line). 
The above expression indicates that the plateau amplitude increases as $\mu(0,T)\approx  \frac{2}{3}(D+D_A)T^2$ when $T\to\infty$. 
This scaling behavior is easily confirmed in Fig.~\ref{Fig3}(a). Eventually, in the infinite-time limit, the PSD becomes the plateau-free form $\mu(f,\infty)\propto f^{-2}$ down to $f\to 0$.

We note that the profile of finite-time PSDs, despite its exact expression~\eqref{Eqn:SingleFreeAOUPPeriodogram} being complex, resembles a Lorentzian form on the whole. Then, the crossover frequency $f_0$ separating the low-$f$ plateau from the high-frequency $f^{-2}$ regime can be well approximated by a Lorentzian interpolation that matches the low- and high-$f$ limits,
\begin{equation}
\mu(f,T)\approx \frac{\mu(0,T)}{1+(f/f_0)^2}.
\label{LorentzianApprox}
\end{equation}
Matching the high-$f$ amplitude to the high-$f$ expansion of the exact form yields an explicit expression for $f_0$, given in Eq.~\eqref{eq:crossoverfreqFreeAOUPPSD}.  Refer to Appendix~\ref{appendix:crossoverfreqfreeAOUP} for derivation and additional analysis. In a nutshell, we obtain $f_0 ^2 \simeq \frac{6}{T^2}$ in the weak-activity limit $ D_A \ll D$ and $f_0 ^2 \simeq \frac{3}{T^2}$ in the strong-activity limit $D_A \gg D$. Thus, in all cases, we confirm that the crossover frequency scales as $f_0 \sim T^{-1}$, in accordance with our physical expectation. This scaling relation is observed in Fig.~\ref{Fig3}(a). 

As clearly visible in Fig.~\ref{Fig3}(b), $\mu(f,T)$ displays an oscillating $f^{-2}$ decay.  Expanding the exact expression at large $f$ yields
\begin{equation}
\begin{aligned}
    \mu(f,T)  
    \approx & \frac{4 D}{f^2 }
    +\frac{2 D_A
   e^{-\frac{T}{\tau_A}} \left(e^{T/\tau_A}
   (T-\tau_A)+\tau_A\right)}{T f^2}
   \\ & -\frac{ 4 k_B \mathcal{T}
   + 2 \zeta D_A \left(1 
    - e^{-\frac{T}{\tau_A}}  \right)}{f^3
   \zeta  T}  \sin \qty(fT) 
   \\ &+ \mathcal{O}(f^{-4}T^{-1}).
\end{aligned}   
\label{eq:SinffreeAOUPPSD}
\end{equation}
In this expression, the first two terms are identified to the dominant $f^{-2}$ contribution, which becomes the infinite-time PSD~\eqref{activeFreePSD} as $T\to\infty$.
Intriguingly, we find that the next order term contains an oscillatory component, which is proportional to $\sin(fT)$. The period of this oscillation is proportional to $T^{-1}$, so the oscillating feature is observable for a short trajectory. Additionally, we notice that the amplitude of this term decays as $1/(f^{3}T)$. Because of this, this oscillation rapidly decays out in the large-$f$ limit. When $T\to\infty$, the oscillatory component decays out in the order of $T^{-1}$.
In summary, $\mu(f,T)$ exhibits finite-time effects at both low and high frequencies, converging to the infinite-time PSD with the convergence speed at a rate of $T^{-1}$.

\section{Power spectral densities for confined AOUPs\label{sec5}}
Now we move to the PSDs of AOUPs confined to a harmonic potential [Eq.~\eqref{eq:AOUP_harmonic_x}]. Contrary to the free AOUP in the previous section, the motion of confined AOUPs is a stationary process, revealing much more complicated PSD patterns, depending on the persistence time $\tau_A$ and  the trap relaxation time $\tau_R$. Moreover, the behaviors of PSDs are closely connected with those of MSDs. In this section, we provide a case-by-case study for all possible distinct patterns of PSDs and conjugated MSDs in a complementary picture.

\subsection{{Infinite-$T$} PSD and MSD \label{secPSDandMSD}}

\begin{figure*}
    \includegraphics[width=0.98\textwidth]{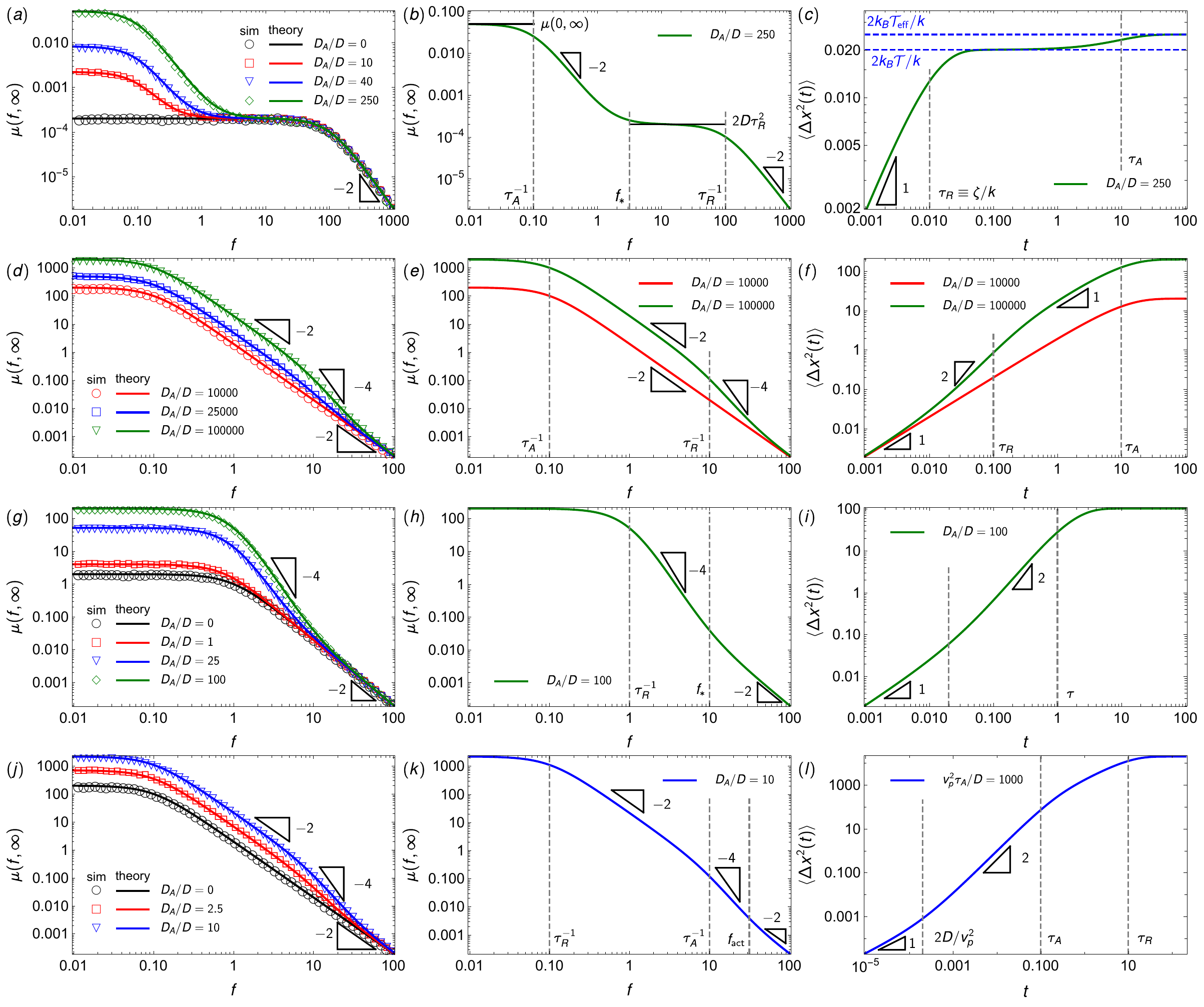}
    \caption{\label{Fig4}
         {Infinite-$T$} PSD and mean-square displacement of AOUPs in a harmonic potential for different parameter regimes.
(a–c) $\tau_A \gg \tau_R$ with weak-to-moderate activity, $v_p < v_p^{*}$ (here $\tau_A = 10$, $\tau_R = 0.01$, and $v_p^{*}$ is the activity scale at which the intermediate thermal plateau in the PSD becomes invisible. See main text for the definition).
(a) PSD of confined AOUPs: symbols show Langevin simulations and solid lines are from Eq.~\eqref{psd_harmonic_compact}.
(b) The prototype PSD for the same parameters as in (a); the theoretical curve is plotted alone to highlight the double-plateau form set by $\tau_R$ and $\tau_A$.
(c) MSD for the same parameters, illustrating the two-plateau structure.
(d–f) Regime $\tau_A \gg \tau_R$ with strong activity, $v_p \gtrsim v_p^{*}$ (again $\tau_A = 10$, $\tau_R = 0.01$).
(d) PSD with simulation data (symbols) and theory (lines).
(e) The prototype PSD for the same regime, showing the theory curve alone for $v_p = v_p ^*$ and $v_p > v_p ^*$.
(f) MSD for the same parameters.
(g–i) In the case of $\tau_A = \tau_R = 1$.
(g) PSD from simulations (symbols) and theory (lines), where the active contribution produces a squared-Lorentzian high-frequency tail.
(h) prototype PSD for the parameters used in (g).
(i) MSD of AOUPs for $\tau_A = \tau_R$.
(j–l) $\tau_A \ll \tau_R$ (here $\tau_A = 0.1$, $\tau_R = 10$).
(j) PSD of confined AOUPs with simulations (symbols) and theory (lines).
(k) prototype PSD for $\tau_A \ll \tau_R$.
(l) MSD for the same parameters.
}
\end{figure*}

The general closed-form expression for the PSD can be analytically obtained using the autocovariance $C_x(t-t')$ of the trapped AOUP in Eq.~\eqref{activeAutocorrelationFtn}, which reads
\begin{align}
& \begin{aligned}
\mu(f,\infty)
&=
 \frac{2D}{{f}^2 + \tfrac{1}{\tau_R ^2}} 
 \\ &
+ \frac{ 2\tau _A  \tau _R^2 v_p ^2 }{\left(\tau _A-\tau _R\right) \left(\tau _A+\tau _R\right)}
\left[ \frac{1}{{f}^2+\tfrac{1}{\tau_A ^2}} - \frac{1}{{f}^2+\tfrac{1}{\tau_R ^2}} \right]
\end{aligned}
\label{ActivePSDInfiniteTime1}
\\ & 
\hphantom{\mu(f,\infty)}
= \frac{2D}{{f}^2 + \tfrac{1}{\tau_R ^2}} 
+
\frac{2 v_p ^2 \tfrac{1}{\tau_A } }{ \qty(f^2 + \tfrac{1}{\tau_A ^2})
\qty(f^2 + \tfrac{1}{\tau_R ^2})
 }
\label{psd_harmonic_compact}.
\end{align}
Equation~\eqref{psd_harmonic_compact} tells us that the power spectrum is the superposition of the well-known Lorentzian function from a trapped thermal Brownian motion and the active contribution
\begin{equation}
    \mu_{\mathrm{act}}=\frac{2 v_p ^2 \tfrac{1}{\tau_A } }{ \qty(f^2 + \tfrac{1}{\tau_A ^2})\qty(f^2 + \tfrac{1}{\tau_R ^2})}.
\end{equation} 
Notably, the active contribution is composed of two Lorentzian functions of the same structure with characteristic frequencies $\tau_A^{-1}$ and $\tau_R^{-1}$ [Eq.~\eqref{ActivePSDInfiniteTime1}].  
In Fig.~\ref{Fig4}, we plot the four distinct patterns of $\mu(f,\infty)$ above, along with the simulated PSDs. The last column shows the MSD~\eqref{Eq:HarmonicMSD} at the same parameters. Before clarifying each case of the PSD, we explain two general properties of the PSD of the stationary active process below.

(1) The Lorentzian form of the PSD indicates that the PSD always entails a zero-frequency plateau, which is completely different from the plateau observed in the PSD of free AOUPs due to finite observation time.  For confined AOUPs,  the zero-frequency value of the spectrum is set by the integral of the autocovariance function
\begin{align}
    \mu({0, \infty}) & = 2\int_0 ^\infty C_x \qty(t) dt= 2\langle x^2\rangle\,\tau
    \label{eq:correlationArea}
\end{align}
with 
\begin{equation}
\tau
=\frac{(D+D_A)\,\tau_R(\tau_A+\tau_R)}{D(\tau_A+\tau_R)+D_A\tau_R},
\label{eq:taucorr_explicit}
\end{equation}
which is the characteristic time of the position correlation~\cite{dechant2023thermodynamic}. Since the long-time MSD plateau is purely a variance, $\langle\Delta x^2(\infty)\rangle=2\langle x^2\rangle$, the above relation states that the zero-frequency plateau is attributed to the long-time plateau in the MSD. 

(2) The PSD has a tail of $\sim f^{-4}$ at high frequencies because of its active component $\mu_{\mathrm{act}}$. Expanding the spectral density~\eqref{psd_harmonic_compact} at $f\to\infty$ yields 
\begin{equation}
\mu(f,\infty) \approx \frac{2D}{f^2 + \tau_R ^{-2}} + \frac{2D_A}{ \tau_A ^2 f^4} +\mathcal{O}(f^{-6}).
\label{eq:highf_expand_main}
\end{equation}
At high frequencies, therefore, the PSD is dominated by the Lorentzian spectrum of confined Brownian motion, followed by the next-order correction term involving $f^{-4}$. Note that the activity-induced $f^{-4}$ scaling in the PSD is absent in the case of free AOUPs (see Eq.~\eqref{PSDfreeAOUPlargef}). In the time domain, consistent with the expansion~\eqref{eq:highf_expand_main},  the MSD is approximated to
\begin{equation}
\langle\Delta x^2(t)\rangle \approx  2D\tau_R \qty(1-e^{-t\slash \tau_R}) +  \frac{v_p^2\tau_R}{\tau_A+\tau_R}  t^2 +\mathcal{O}(t^3).
\label{eq:smallt_expand_main}
\end{equation}
Thus, for the confined AOUP, the activity produces a fast $f^{-4}$ tail at high frequencies and transient ballistic MSD in the short time. 
This shows that  the transient ballistic term is indeed responsible for the tail of $f^{-4}$ at high frequencies.  
Having these information in mind, in the following, we investigate the four distinct cases of the spectral densities highlighted in Fig.~\ref{Fig4}.

\subsubsection{Long-Persistence Regime ($\tau_A\gg \tau_R$) and Moderate Activity ($\sqrt{D/\tau_A}<v_p<v_p^*$)}
Consider the power spectra under the condition of $\tau_A\gg\tau_R$. In this situation, the persistence frequency $\tau_A^{-1}$ and the trap-relaxation frequency $\tau_R^{-1}$ are well separated. Remarkably, as shown in Fig.~\ref{Fig4}(a)--(c), the power spectra and MSDs exhibit two plateaus when the activity strength is not too strong. In the opposite case of strong activities, the two-plateau features are masked (Fig.~\ref{Fig4}(d)--(f)). Here, we start with the PSD showing the two-plateau profile (at the moderate activity).  


With a long persistence time ($\tau_R\ll\tau_A$) and a proper propulsion speed, the confined active particle reveals the so-called double trapping dynamics~\cite{kim2024active}. 
The physical picture is as follows. First, the particle's thermal agitation in the harmonic trap relaxes on the timescale of $\tau_R$ during which the active force is effectively quasi-static. Second, on the timescale $\tau_A$, the active displacements are decorrelated and their relaxation dynamics are observed. This phenomenon was reported in our earlier simulation study of diffusion of active Brownian particles in a polymer network~\cite{kim2024active}. This two-stage confinement is shown in the MSD curve in Fig.~\ref{Fig4}(c). 
The MSD  has the thermal plateau $2k_B\mathcal{T}/k$ for $\tau_R\ll t\ll\tau_A$ and eventually saturates to the second plateau at $2k_B\mathcal{T}_{\mathrm{eff}}/k$ for $t\gtrsim\tau_A$.

This double trapping also results in a double-plateau profile of  the PSDs (Fig.~\ref{Fig4}(a) and \ref{Fig4}(b)). 
In the intermediate frequency window of $\tau_A^{-1}\ll f\ll\tau_R^{-1}$, the PSD exhibits a plateau relating to the thermal relaxation in a harmonic potential
\begin{equation}
\mu(f,\infty)\simeq 2D\tau_R^2 =
 2\left(\frac{k_B\mathcal{T}}{k}\right)\tau_R.
\label{eq:thermal_plateau_longpersist_main}
\end{equation}
This is precisely the PSD property explained in Eqs.~\eqref{eq:correlationArea}~and~\eqref{eq:taucorr_explicit}. Equation~\eqref{eq:thermal_plateau_longpersist_main} can be understood as the product of the thermal plateau and its relaxation time $\tau_R$. 
In the low frequency regime $f\ll\tau_A^{-1}$, as noted in Eq.~\eqref{eq:correlationArea} the PSD reaches the second plateau
\begin{equation}
\mu(f,\infty)=
 2\left(\frac{k_B\mathcal{T}_\mathrm{eff}}{k}\right)\tau.
\end{equation}
with the characteristic time $\tau$ defined in Eq.~\eqref{eq:taucorr_explicit}.

Now we quantify the condition that the double plateau (i.e., double trapping) is clearly visible. (i) The active displacement should be appreciably larger than the thermal displacement, i.e., $D_A\gg D$, which in turn leads to the first condition
\begin{equation}
v_p\gg\sqrt{D/\tau_A}.
\end{equation}
(ii) The intermediate thermal PSD plateau should not be masked by the active component of the PSD, meaning that the activity is not too strong. This condition can be found in the following. Let us define a crossover frequency $f_*$ within the intermediate window ($\tau_A^{-1}\ll f\ll\tau_R^{-1}$) as the frequency at which the PSD's active contribution 
\begin{equation}
\mu_{\mathrm{act}}(f,\infty)\simeq \frac{2v_p^2\tau_R^2}{\tau_A f^2}
\end{equation}
becomes comparable to the thermal plateau. Equating $\mu_{\mathrm{act}}(f_*,\infty)\approx 2D\tau_R^2$ yields
\begin{equation}
f_* \approx \tau_A ^{-1}\sqrt{\frac{D_A }{D}}.
\label{eq:fstar}
\end{equation}
For a double-plateau PSD, the requirement is $f_*\ll\tau_R^{-1}$ (see Fig.~\ref{Fig4}(b)), which sets the second condition
\begin{equation}
v_p\ll \frac{\sqrt{D\tau_A}}{\tau_R}\equiv v_p^*.
\end{equation}
Taken together, we find that the power spectrum has the double plateau structure when the activity strength lies in $\sqrt{D/\tau_A}\ll v_p\ll v_p^*$.

\subsubsection{Long-Persistence Regime ($\tau_A\gg \tau_R$) and Strong Activity}

When the activity is substantially strong such that $v_p\gtrsim v_p^*$, the crossover frequency has $f_*\gtrsim\tau_R^{-1}$. In this case, albeit $\tau_A\gg\tau_R$, the PSD loses the intermediate thermal plateau while keeping the zero-frequency plateau. See the simulated PSDs (Fig.~\ref{Fig4}(d)) and theoretical curves (Fig.~\ref{Fig4}(e)). 
Here, in the intermediate frequency window, the PSD exhibits the scaling feature resulting from the active Fickian diffusion: 
\begin{equation}
\mu(f,\infty)\simeq \frac{2D_{\mathrm{mid}}}{f^2},
\qquad
D_{\mathrm{mid}}=D_A \qty(\frac{\tau_R}{\tau_A})^2. 
\label{eq:Dmid_longpersist_main}
\end{equation}
Concurrently, the MSD in Fig.~\ref{Fig4}(f) exhibits the Fickian regime 
$\langle\Delta x^2(t)\rangle\simeq 2D_{\mathrm{mid}}t$ for $\tau_R\ll t\ll\tau_A$, before the saturation at $2k_B\mathcal{T}_{\mathrm{eff}}/k$.

When $f \gg  \tau_R ^{-1} $, the thermal (Lorentzian) and active components  of the PSD~\eqref{psd_harmonic_compact} behave as $\mu_\mathrm{th}\simeq2D\slash f^2$ and $\mu_\mathrm{act} \simeq 2v_p ^2 \tau_A ^{-1} \slash f^4$, respectively. In the frequency range of $\tau_R ^{-1} \ll f \ll f_{*}$, it can be shown that $\mu_\mathrm{act}\gg \mu_\mathrm{th}$, leading to the $f^{-4}$ scaling in the PSD. This is illustrated in Fig.~\ref{Fig4}(d) and \ref{Fig4}(e). As pointed out earlier, the $f^{-4}$ scaling in the PSD is attributed to the activity-originated ballistic scaling $\sim t^2$ in the MSD [Fig.~\ref{Fig4}(f)]. 
When $f\gg f_*$, the PSD is dominated by the thermal Lorentzian part alone, thus $\mu(f,\infty)\simeq 2D/f^2$. The corresponding regime in the MSD is $\langle \Delta x^2 (t) \rangle \simeq 2Dt$ for $t \ll 2D \slash v_p ^2 $.

\subsubsection{Cross-over Persistence: $\tau_A=\tau_R$}
In the case of $\tau_A=\tau_R$, the PSD~\eqref{ActivePSDInfiniteTime1} has the special form 
\begin{equation}
    \mu(f,T=\infty) = \frac{2D}{{f}^2 + \tau_{R} ^{-2}} +
\frac{2D_A }{ \tau_{R} ^2 \qty(f^2 + {\tau_{R} ^{-2}})^2 
 }.
 \label{eq:PSD_tauAeqtauR}
\end{equation}
See the profile of the PSD in Fig.~\ref{Fig4}(g) and \ref{Fig4}(h). Under this condition, the functional form of the active part $\mu_\mathrm{act}$, originally a sum of two Lorentzians, is changed to a single squared Lorentzian. This implies that the time-domain positional autocorrelation is no longer a sum of exponentials. 
Instead, it contains a polynomial prefactor multiplying the exponential decay
\begin{equation}
C_{x}(t)
=
D\tau_{R}\,e^{-t/\tau_{R}}
+
\frac{D_A\tau_{R}}{2}\qty(1+\frac{t}{\tau_{R}} )e^{-t/\tau_{R}}.
\label{eq:Cxx_tauAeqtauR_final}
\end{equation}
The MSD is $\langle\Delta x^2(t) \rangle = 2C_x(0) - 2C_x(t)$, which reads
\begin{equation}
\begin{aligned}    
    \langle \Delta x^2 (t) \rangle & =
    2D \tau_{R} \qty(1 - e^{- t\slash \tau_{R}}) + D_A \qty[\tau_{R} - \qty(\tau_{R}+t)e^{-t\slash \tau_{R}}].
\end{aligned}
\label{eq:msdtauAtauR}
\end{equation}

In the cross-over persistence case, the PSD exhibits a unique pattern, as shown in Fig.~\ref{Fig4}(g)~and~\ref{Fig4}(h). In Eq.~\eqref{eq:PSD_tauAeqtauR}, both the thermal Lorentzian and active terms crossover at the same frequency, $\tau_R ^{-1}$. As a result, the PSD crosses over directly from the zero-frequency plateau to the slope of $2D_A \tau_R ^{-2}\slash f^4$, without the intermediate $f^{-2}$-scaling regime. 
This $f^{-4}$ scaling originates from  a transient ballistic dynamics as noted in Eq.~\eqref{eq:highf_expand_main} and is visible if $D_A \gg D \qty(f \tau_{R})^2 $. In accordance with this behavior, the MSD~\eqref{eq:msdtauAtauR} exhibits $\sim t^2$ scaling on the corresponding time scale (Fig.~\ref{Fig4}(i)). 

\subsubsection{Short-Persistence Regime ($\tau_A \ll \tau_R $)\label{sec5A4}}
Under the short-persistence condition $\tau_A\ll\tau_R$, the trap-induced relaxation dynamics is much slower than the correlation time of the active force.  
As a result, confinement is irrelevant over a broad range of lag times, so the MSD resembles that of a free AOUP before the final saturation set by the trap; see the MSD curve in Fig.~\ref{Fig4}(l). Here, the MSD has the following four distinct dynamic regimes: 
\begin{equation}
    \begin{aligned}
        \langle \Delta x^2 (t) \rangle \simeq 
        \begin{cases}
            2Dt & t \ll 2D / v_p ^2\\
            v_p^2 t^2 & 2D / v_p ^2 \ll t \ll \tau_A\\
            2D_{\mathrm{eff}}t & \tau_A \ll t \ll \tau_R \\
            2k_B \mathcal{T}_{\mathrm{eff}} / k & t \gg \tau_R 
        \end{cases}\;.
    \end{aligned}
\label{eq:MSD_fastpersist_regimes}
\end{equation}
In particular, in the intermediate-time window $\tau_A\ll t\ll\tau_R$, the MSD is Fickian with the slope $2D_{\mathrm{eff}}$, since the active force has already decorrelated while the particle has not yet felt the trap.

Correspondingly, the PSD displays the four frequency responses (Fig.~\ref{Fig4}(j)--(k)). 
At a very low frequency $f\ll\tau_R^{-1}$ the spectral density reaches the confinement plateau $\mu(0,\infty)=2\langle x^2\rangle\tau$.
In the subsequent (broad) intermediate range $\tau_R^{-1}\ll f\ll\tau_A^{-1}$, confinement is negligible on the corresponding timescales, while the active force is already decorrelated; thus the PSD follows the free-diffusion scaling
\begin{equation}
\mu(f,\infty)\simeq \frac{2D_{\mathrm{eff}}}{f^2}.
\label{eq:PSD_mid_fastpersist_main}
\end{equation}
At high frequencies of $\tau_A^{-1} \ll f\ll v_p^2/2D$, the PSD exhibits the activity-induced $f^{-4}$ scaling attributed to the ballistic diffusion $v_p^2 t^2$ at short times. Beyond $f\approx v_p^2/2D$, the thermal agitation is solely dominating in the dynamics of confined active entities, resulting in $\mu\simeq 2D/f^2$.

\subsection{{ Finite-$T$} PSD}
\begin{figure}
\includegraphics[width=0.45\textwidth]{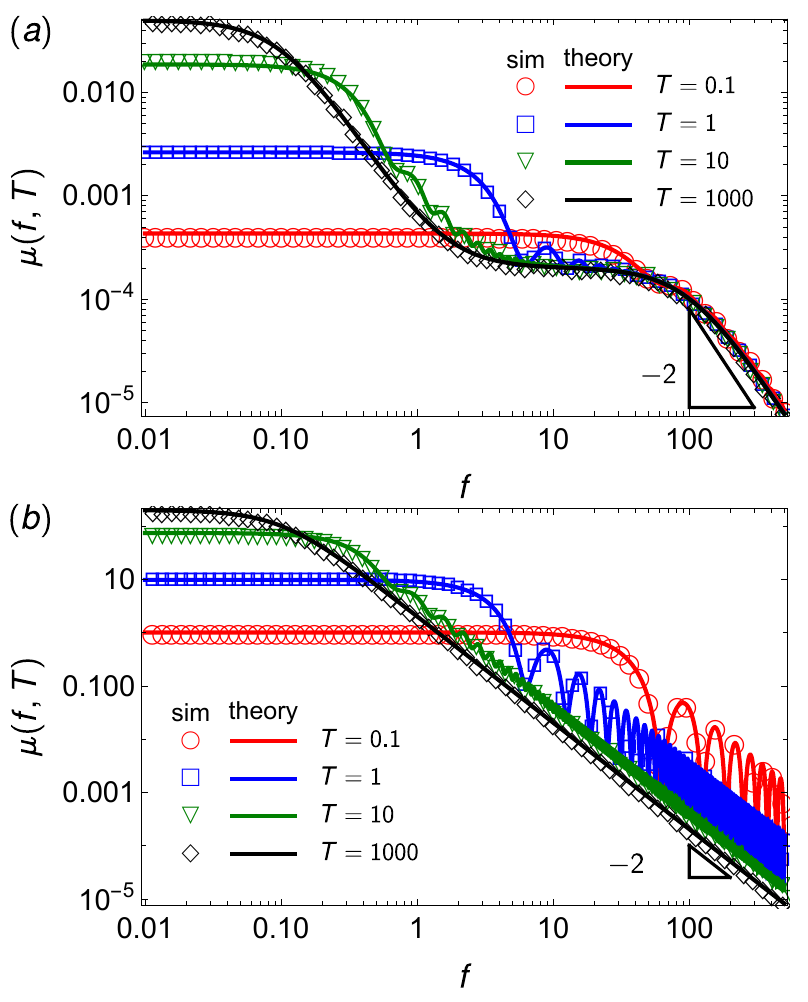}
\caption{\label{Fig5}
{Finite-$T$} PSDs for AOUPs in a harmonic potential. The symbols display the numerical values calculated with the simulated trajectories with $\tau_A = 10$ and $\tau_R = 0.01$, and the solid lines depict the analytical result of the PSD~\eqref{eq:ST_decomp}.
(a) The double-plateau case, corresponding to long-persistence time $\tau_A\gg\tau_R$ and moderate activity $v_p \ll v_p^*$, at $D_A/D=250$. 
(b) The cross-over case $v_p=v_p^*$. 
The oscillatory component is clearly resolved and persists up to higher frequencies.
}
\end{figure}

We finally turn to the finite-$T$ PSD $\mu(f,T)$ for confined AOUPs. The PSD~\eqref{eq:finitePSDReal} can be evaluated analytically using the autocovariance~\eqref{activeAutocorrelationFtn}. It turns out that the power spectrum decomposes into three contributions,
\begin{equation}
  \mu (f,T) 
  = \mu(f,\infty)
  + \mu_{\mathrm{nos}}(f,T)
  + \mu_{\mathrm{os}}(f,T),
  \label{eq:ST_decomp}
\end{equation}
where $\mu(f,\infty)$ is the infinite-$T$ PSD~\eqref{psd_harmonic_compact},
$\mu_{\mathrm{nos}}(f,T)$ is a non-oscillatory finite-time contribution, and
$\mu_{\mathrm{os}}(f,T)$ is an oscillatory finite-time contribution  containing $\cos(fT)$ and $\sin(fT)$ terms.
The exact analytic expressions for $\mu_{\mathrm{nos}}(f,T)$ and $\mu_{\mathrm{os}}(f,T)$ are provided in Eqs.~\eqref{eq:Delta_S_nonosc_appendix}~and~\eqref{eq:S_osc_appendix} of Appendix~\ref{exactperiodogramConfined}.

In Fig.~\ref{Fig5}, we  present finite-$T$ PSDs $\mu(f,T)$ obtained from finite-length simulated trajectories, together with the analytic expression~\eqref{eq:ST_decomp}. Panel (a) presents the regime $\tau_A \gg \tau_R$ with moderate activity $v_p \ll v_p^*$, where the infinite-time PSD exhibits the double-plateau structure. Panel (b) shows the cross-over case of $v_p = v_p^*$.
The remaining two cases of confined AOUPs introduced in the previous section display qualitatively similar finite-time behaviors to those presented here, and are therefore not shown separately. 

The first noteworthy property is the $T$-dependency in the zero-frequency amplitude. As a general trend, the amplitude decreases monotonically as $T$ decreases.  From the PSD~\eqref{eq:ST_decomp}, the zero-frequency value of the power spectrum is found to be
\begin{equation}
\begin{aligned}
&\mu(0,T)  =   \frac{2 k_B \mathcal{T} \tau_R  }{k}  
\left( 1 + \frac{\tau_R}{T} \left(e^{-T\slash \tau_R }-1\right) \right)
\\ & + 2 D_A \tau_R ^2 
\left( \frac{ \tau_A ^3 \left(e^{-\frac{T}{\tau_A }}-1\right)
-\tau_R ^3 \left(e^{-\frac{T}{\tau_R}}-1\right)
   }{T \left(\tau_A ^2-\tau_R ^2\right)}  +1\right).
\end{aligned}
\end{equation}
In the large-$T$ limit ($T\gg \tau_R,\tau_A$), $\mu(0,T)$ converges to the $T$-independent plateau $\mu=2(k_B\mathcal{T}_\mathrm{eff}/k)\tau$, consistent with Eq.~\eqref{eq:thermal_plateau_longpersist_main}. In the opposite limit of short observation windows, $T\ll\tau_R,\tau_A $, the particle does not have a sufficient time to complete either a full trap relaxation or a full active persistence excursion. In this regime, the zero-frequency amplitude reduces to
\begin{equation}
\begin{aligned}
  \mu (0,T)
  & \simeq \langle x^2\rangle T  = \frac{k_B \mathcal{T}_{\mathrm{eff}}}{k} T .
\end{aligned}
  \label{eq:short_T_flat}
\end{equation}
Hence, in the short-$T$ limit, the zero-frequency plateau grows linearly with $T$, as clearly observed in both panels (a) and (b) of Fig.~\ref{Fig5}. Intriguingly, in this limit, the PSD is proportional to the variance of $x$.
We note that this variance-dependent plateau is a finite-time effect, clearly distinguished from the behavior of the free AOUP, where the zero-frequency amplitude of the finite-$T$ PSD grows as $\mu(f=0)\propto T^{2}$. 

The second notable feature is the pronounced oscillation at high frequencies, as shown in Fig.~\ref{Fig5}(b). The oscillation amplitude increases as the trajectory length is shorter. Notably, in contrast to the oscillatory behavior of $\mu(f,T)$ for free AOUPs, the PSD oscillations in confined AOUPs do not decay at high frequencies.  

To comprehend this behavior, we analyze the two finite-time contributions, $ \mu_{\mathrm{nos}}(f,T)$ and $ \mu_{\mathrm{os}}(f,T)$ in Eq.~\eqref{eq:ST_decomp}. From the exact expression Eq.~\eqref{eq:Delta_S_nonosc_appendix}, the non-oscillating term behaves as
\begin{equation}
  \mu_{\mathrm{nos}}(f,T)\sim\mathcal{O}\left({ f^{-2}}T^{-1}\right)
  \label{eq:nonosc_scaling_main}
\end{equation}
at high frequencies or in the large-$T$ limit. The finite-time correction therefore vanishes inversely with the trajectory length $T$, while its frequency dependence follows the same $f^{-2}$ decay as the infinite-time PSD, as shown in Eq.~\eqref{eq:PSD_mid_fastpersist_main}. 

For high frequencies ($f\gg\tau_R^{-1},\tau_A ^{-1}$), the oscillatory contribution is approximated as
\begin{equation}
\begin{aligned}
  \mu_{\mathrm{os}}(f,T)
  & \simeq -A(f,T)\cos(fT) + \mathcal{O} \qty(f^{-3} T^{-1}),
  \label{eq:Sosc_main}
\end{aligned}
\end{equation}
where the leading amplitude $A(f,T)$ is 
\begin{equation}
\begin{aligned}
  A(f,T) 
  & \equiv \frac{2 D e^{-T\slash \tau_R}  \tau_R}{f^2  T}
+ 
  \frac{2 D_A \tau _R^2 \left(\tau _A e^{-\frac{T}{\tau _A}}-\tau _R e^{-\frac{T}{\tau _R}}\right)}{f^2 T \left(\tau _A^2-\tau _R^2\right)}
  \label{eq57}
  \\ & 
  = e^{-T/\tau_R} \mathcal{O} (f^{-2}T^{-1})+  e^{-T/\tau_A} \mathcal{O} (f^{-2}T^{-1}).
\end{aligned}
\end{equation}
Thus, the oscillatory term's magnitude decays as $A(f,T)\sim (T f^{2})^{-1}$, modulated by the exponential suppression factors $e^{-T/\tau_R}$ and $e^{-T/\tau_A}$. This analysis demonstrates that  $\mu_{\mathrm{os}}(f,T)$ decays with the same $T$-dependence as the non-oscillatory term $\mu_\mathrm{nos}$, and its high-frequency dependence also follows $f^{-2}$, matching that of the infinite-time PSD. Consequently, the oscillatory component remains relevant over a broader frequency range for confined AOUPs.  This behavior contrasts markedly with the free AOUP model, in which the oscillatory term decays as $f^{-3}$, faster than the decay of the infinite-time PSD, accordingly becoming negligible as $f\to\infty$.

\section{Discussion and Conclusions\label{sec6}}

{In this work, we have analytically studied the power spectral densities of trajectories of active Ornstein–Uhlenbeck particles, both in free space and under harmonic confinement, under experimentally relevant conditions where trajectories are recorded over a finite observation time $T$. We have derived explicit finite-$T$ expressions and determined their infinite-$T$ limits, corresponding to the conventionally defined power spectral densities, which had remained unknown previously.}

In free space, where the particle position is nonstationary, the infinite-$T$ PSD retains the Brownian noise spectrum $\mu(f,\infty)\propto f^{-2}$ over the entire frequency range. Notably, activity does not introduce a new spectral exponent (e.g., $f^{-4}$); instead, it appears through a Lorentzian correction that merely alters the prefactor of the $f^{-2}$ spectrum across the crossover frequency $f\sim \tau_A^{-1}$. This structure contrasts with the MSD at short times, where the thermal displacement is dominant and the effect of the activity-induced ballistic movement remains subleading, whereas the PSD at high frequencies reveals an activity-dependent amplitude. 

When the trajectory is limited by a finite observation window, the resulting finite-time PSD contains a window-dependent low-frequency plateau, of which amplitude increases as $T^2$ for large $T$. At high frequencies, the finite-time PSD has an oscillation on the $f^{-2}$ tail, with an oscillation magnitude decaying as $f^{-3}$.

Under harmonic confinement, the AOUP dynamics is stationary, and the PSD is closely related to the MSD. In this case, the PSD consists of the well-known Lorentzian spectrum of a trapped Brownian particle and the sum of two Lorentzian contributions arising from active noise and harmonic confinement with the persistence time $\tau_A$ and the trap relaxation time $\tau_R$, respectively. Based on this structure, we have identified the spectral features of confined AOUPs governed by the interplay between $\tau_A$, $\tau_R$, and the activity strength Pe. Depending on the parameter regime, the PSD exhibits several distinct spectral profiles that are not observed in thermal particles or free AOUPs. When $\tau_A\gg \tau_R$ with a moderate activity, $\sqrt{D\slash \tau_A} \ll v_p \ll v_p^*$, a double-trapping mechanism emerges, leading to a characteristic double-plateau profile in the PSD. For strong activities, $v_p \gtrsim v_p ^*$, the intermediate plateau is masked, and the PSD structure exhibits both $f^{-2}$ and $f^{-4}$ intermediate scalings. In the special case $\tau_A=\tau_R$, the PSD reduces to a squared-Lorentzian and displays only the intermediate $f^{-4}$ scaling. 
For $\tau_A \ll \tau_R$, both intermediate $f^{-2}$ and $f^{-4}$ scalings appear again, however, originating from a distinct physical mechanism. 
Overall, confined AOUPs exhibit a range of qualitatively different spectral structures, including double-plateau profiles and the new $f^{-4}$ scaling, depending on the underlying dynamical regime. These features are absent in both thermal Brownian systems and free active diffusion.  Notably, the $f^{-4}$ scaling is connected to transient ballistic  MSD of $\sim t^2$. 

Finally, we have investigated finite-window effects on {the PSD of the trajectories of confined AOUPs}. We have shown that the short-time spectrum develops a plateau whose value is proportional to $T$, which is distinct from the low-frequency confinement plateau of the stationary PSD. The finite-time PSD of confined active systems also contains an oscillatory component, but unlike the free active system, where the oscillation decays rapidly as $f^{-3}$, here it persists with an $f^{-2}$ decay. Our finite-$T$ theory therefore provides an analytical framework for understanding how finite observation windows modify the measured PSD and what additional structures they generate.

Our study provides an exactly solvable reference for interpreting the stochastic trajectories of active systems. Additionally, we discuss how the theoretical results obtained in this work can provide quantitative insight into active dynamics in real experiments.

An interesting example is the active-matter experiment reported in Ref.~\cite{maggi2014generalized}, which studies a colloidal particle immersed in a bacterial bath in a cylindrical vessel.
In that setting, the motion along the $y$ direction can be modeled as a harmonically confined AOUP.
The measured MSD exhibits a clear active contribution beyond the thermal plateau, indicating that nonequilibrium forces cannot be neglected.
Under the reported experimental conditions, the active correlation time is approximately $0.093\,\mathrm{s}$, whereas the relaxation time associated with the harmonic confinement is approximately $3.46\,\mathrm{s}$.
The system therefore lies in the regime $\tau_A \ll \tau_R$ of our theory. Accordingly, our analysis predicts an MSD of the type shown in Fig.~\ref{Fig4}(l), and Ref.~\cite{maggi2014generalized} indeed displays a qualitatively similar behavior. 

Another relevant example is an experimental study of the optically trapped Janus colloid  with a simulation of harmonically bound active Brownian particles~\cite{halder2025interplay}.
Although their system was made with active Brownian particles, the resulting MSDs and PSDs can still be understood within our confined AOUP model. The confined Janus particle exhibits the MSD patterns similar to the ones in Fig.~\ref{Fig4}(f) in our study. 
Notably, in their Fig.~6(c), the authors presented the PSD from the simulation of active Brownian particles with the ratio of $\tau_A/\tau_R=0.01$, reporting that the PSD deviates from the $f^{-2}$-Lorentzian spectrum of free Brownian motion. Beyond this description, our study provides further quantitative information on the obtained PSD. Their simulated system belongs to the short-persistence regime ($\tau_A\ll \tau_R$) studied in Sec.~\ref{sec5A4}. As illustrated in Fig.~\ref{Fig4}(j), in this case, the PSD displays the three distinct scaling regimes ($f^{-2}\to f^{-4}\to f^{-2}$) as the frequency decreases from infinity to zero, before approaching the zero-frequency plateau. Indeed, such a scaling variation is visible in their PSD curve.

Beyond such relatively simple systems, the current work can also help interpret more complex active biological systems. For instance, in Ref.~\cite{narinder2026time}, the authors measured the AFM-tip position in the actin cortex and tested the fluctuation--dissipation theorem by comparing the renormalized PSD with the dissipative response. At equilibrium, these two quantities coincide, whereas the experimental data reveal a mismatch at low frequencies. This observation is understandable with our confined AOUP framework, where the nonequilibrium fluctuations play a crucial role in the intermediate- or low-frequency domains of the PSD (and typically increase the amplitude of the low-frequency part of the PSD), while thermal fluctuations dominate in the high-frequency regime.

Several directions remain open for future study.
One important extension is to construct and analyze more realistic models that incorporate viscoelasticity and non-Markovian environments~\cite{joo2023viscoelastic}.
Another is to go beyond the first moment of the single-trajectory PSD and investigate higher-order statistics, in particular the second moment or coefficient of variation.
The present work already contributes to a better understanding of active-particle trajectories, and, when combined with these further developments, it can serve as a useful starting point for a broader frequency-domain theory of active motion.

\appendix

\section{Numerical simulations\label{label:sim}}

The numerical simulation of our Langevin equation models, i.e.,  Eqs.~\eqref{FreeAOUPLangevin} \& \eqref{AOUPactiveForceLangevin} for free AOUPs and  Eqs.~\eqref{eq:AOUP_harmonic_x} \& \eqref{AOUPactiveForceLangevin} for AOUPs in a harmonic potential, was implemented using the Euler--Maruyama scheme with a fixed time step $\Delta t = 5\times 10^{-5}$.
For each parameter set, we generated $N_{\mathrm{traj}}=1000$ independent trajectories.

For the free AOUP,
\begin{align}
x_{n+1} &= x_n + \frac{\Delta t}{\zeta}\Bigl(\xi_n + F_n\Bigr),\label{eq:app_free_update_x}\\
F_{n+1} &= F_n - \frac{\Delta t}{\tau_A}F_n
+ \sqrt{\frac{2\zeta^2 v_p^2}{\tau_A}\Delta t}\;\eta_n,\label{eq:app_update_F}
\end{align}
and for the harmonically confined AOUP,
\begin{equation}
x_{n+1} = x_n + \frac{\Delta t}{\zeta}\Bigl(-k x_n + \xi_n + F_n\Bigr),
\label{eq:app_conf_update_x}
\end{equation}
together with Eq.~\eqref{eq:app_update_F}, where $\eta_n$ and $\xi_n$ are independent standard Gaussian variables, and the thermal noise increment is
$\xi_n \Delta t = \sqrt{2\zeta k_BT\,\Delta t}\;g_n$ with $g_n\sim\mathcal{N}(0,1)$. 
For the free AOUP, we set $x_0=0$ and sampled $F_0$ from the stationary Gaussian distribution of the active OU process.
In order to implement the stationary initial condition assumed in the theory, in the case of harmonic confinement, we sampled $(x_0,F_0)$ from the stationary joint Gaussian distribution computed from the exact stationary second moments.

For a discrete trajectory $\{x_n\}_{n=0}^{N-1}$ with $t_n=n\Delta t$ and $T=N\Delta t$, we approximated
\begin{equation}
\int_{0}^{T} x(t)e^{-ift}\,dt \;\approx\; \Delta t\sum_{n=0}^{N-1} x_n\,e^{-if t_n}.
\label{eq:app_riemann}
\end{equation}
Ensemble-averaged spectra $\mu(f,T)$ were obtained by averaging over $N_{\mathrm{traj}}$ trajectories.

\section{Free AOUPs}

\subsection{Exact closed-form expression for {finite-$T$} PSD\label{sec:freeperiodogram}}
Here we provide the finite-$T$ PSD of free AOUPs observed over a finite time window $T$. 
For a trajectory defined in the interval of $0\le t\le T$, using the two-time covariance in Eq.~\eqref{CxxFreeAOUP} together with Eq.~\eqref{eq:finitePSDReal} yields the exact closed-form expression: 
\begin{widetext}
\begin{equation}
    \begin{aligned}
        \mu(f,T) =
\frac{e^{-T/\tau_A}}{f^3 T\, \zeta\, (1 + f^2 \tau_A^2)^2}
& \Big[ 
\; 2 f^2 v_p^2 \zeta \tau_A^3 \Big( 
    f \tau_A 
    + f^3 \tau_A^3 
    - 2 f \tau_A \cos(f T) 
    + ( -1 + f^2 \tau_A^2 ) \sin(f T)
\Big) \\
& \quad +\; f v_p^2 \zeta \tau_A^2 (1 + f^2 \tau_A^2) \cos(f T) 
\\ 
& \quad \quad +\; 2 e^{T/\tau_A} \Big(
    f \Big(
        2 k_B \mathcal{T} T (1 + f^2 \tau_A^2)^2 
        + v_p^2 \zeta \tau_A \Big(
            -\tau_A (1 + f^4 \tau_A^4) 
            + T (2 + 3 f^2 \tau_A^2 + f^4 \tau_A^4)
        \Big)
    \Big) \\
& \quad\quad \quad
    \vphantom{\Big( (1 + f^2 \tau_A^2) (2 k_B T)} 
    -\; (1 + f^2 \tau_A^2) \Big(
        2 k_B \mathcal{T} (1 + f^2 \tau_A^2) 
        + v_p^2 \zeta \tau_A (2 + f^2 \tau_A^2)
    \Big) \sin(f T)
\Big)
\Big] \; .
    \end{aligned}
    \label{Eqn:SingleFreeAOUPPeriodogram}
\end{equation}
\end{widetext}

\subsection{Crossover frequency in the {finite-$T$} PSD\label{appendix:crossoverfreqfreeAOUP}}
Using the Lorentzian approximation as in Eq.~\eqref{LorentzianApprox}, we now find the analytic expression for the crossover frequency $f_0$ in the $\mu(f,T)$ for free AOUPs. 
Based on the Lorentzian interpolation~\eqref{LorentzianApprox}, we define the crossover frequency as follows: 
\begin{align}
    f_0 ^2 & = \frac{\lim_{f\to\infty} \mu(f,T) f^2}{\mu(0,T)},
\end{align}
which yields the analytic expression
\begin{widetext}
    \begin{equation}
        \begin{aligned}
            f_0 ^2 & =  
            \frac{1}{T^2} \frac{12D + 6 D_A  \qty(1 - \frac{\tau_A}{T} + \frac{\tau_A}{T} e^{-T\slash \tau_A})}{
            2D + 2 D_A \qty(1- \frac{3}{2} \frac{\tau_A}{T} - \frac{3\tau_A^2}{T^2}  e^{-T\slash \tau_A} -
            \frac{3 \tau_A ^3}{T^3} e^{-T\slash \tau_A} + \frac{3 \tau_A^3}{T^3} )
            }.
        \end{aligned}
        \label{eq:crossoverfreqFreeAOUPPSD}
    \end{equation}
\end{widetext}
In the limit of $D_A \to 0$ (equivalently, $v_p \to 0$ or $\tau_A\to0$), we obtain $f_0 ^2 \sim 6\slash T^2$. In the opposite limit, $D_A \to \infty$, we get $f_0^2 \sim 3\slash T^2$. 
In both weak- and strong-activity limits we find $f_0\propto T^{-1}$, with a difference in the numerical prefactor.

\section{Confined AOUPs}

\subsection{Exact closed-form expression for {finite-$T$} PSD\label{exactperiodogramConfined}}
Here we provide the analytic expressions for $\mu_\mathrm{nos}(f,T)$ and $\mu_\mathrm{os}(f,T)$ introduced in the expression of finite-$T$ PSD, Eq.~\eqref{eq:ST_decomp}, in the main text, which read
\begin{widetext}
    \begin{align}
        \mu_{\mathrm{nos}}(f,T) & =\frac{2 k_B \mathcal{T} \tau_R ^3 (f^2 \tau_R ^2 -1 )}{\zeta T (1+f^2 \tau_R ^2)^2}  \nonumber
            \\ & 
            \quad +
            \frac{2 \tau_A  \tau_R ^2 v_p ^2}{T \left(f^2 \tau_A ^2+1\right)^2 \left(f^2 \tau_R ^2+1\right)^2 (\tau_A +\tau_R )}
            \left[ 
\vphantom{\frac{v_p ^2}{\left(f^2 \tau_A ^2+1\right)^2 \left(f^2 \tau_R ^2+1\right)^2 (\tau_A +\tau_R )}}         
\tau_A ^4 \left(f^3 \tau_R ^2+f\right)^2 + \tau_A  \tau_R  \left(f^2 \tau_R ^2-1\right) + f^2 \tau_R ^4
\right.
\nonumber
\\ & \qquad \qquad \qquad \qquad \qquad \qquad \qquad \qquad \qquad \qquad 
\left.
\vphantom{\frac{v_p ^2}{\left(f^2 \tau_A ^2+1\right)^2 \left(f^2 \tau_R ^2+1\right)^2 (\tau_A +\tau_R )}}
+\tau_A ^3
   \left(3 f^4 \tau_R ^3+f^2 \tau_R \right)+\tau_A ^2 \left(2 f^4 \tau_R ^4-f^2 \tau_R ^2-1\right)-\tau_R ^2 
            \right]  
            \label{eq:Delta_S_nonosc_appendix}
        \\ 
        \mu_{\mathrm{os}}(f,T) 
        & = -\frac{2 k_B \mathcal{T} \tau_R ^3 \left(f^2 \tau_R ^2-1\right) e^{-\frac{T}{\tau_R }} }{\zeta  T \left(f^2 \tau_R ^2+1\right)^2} \cos (f T) 
        -\frac{4 f k_B \mathcal{T} \tau_R ^4 e^{-\frac{T}{\tau_R }}}{\zeta  T \left(f^2 \tau_R ^2+1\right)^2} \sin (f T)
        \nonumber
        \\ & \quad 
        + \frac{2 \tau_A  \tau_R ^2 v_p ^2 e^{-T \left(\frac{1}{\tau_A }+\frac{1}{\tau_R }\right)} \left(\tau_R ^3 \left(f^2
   \tau_A ^2+1\right)^2 \left(f^2 \tau_R ^2-1\right) e^{T/\tau_A }-\tau_A ^3 \left(f^2 \tau_A ^2-1\right) \left(f^2 \tau_R ^2+1\right)^2 e^{T/\tau_R }\right)}{T \left(f^2 \tau_A ^2+1\right)^2 \left(f^2 \tau_R ^2+1\right)^2 (\tau_A -\tau_R )
   (\tau_A +\tau_R )} \cos (f T)
   \nonumber
   \\ & \quad -
   \frac{4 f \tau_A  \tau_R ^2 v_p ^2 e^{-T \left(\frac{1}{\tau_A }+\frac{1}{\tau_R }\right)} \left(\tau_A ^4 \left(f^2
   \tau_R ^2+1\right)^2 e^{T/\tau_R }-\tau_R ^4 \left(f^2 \tau_A ^2+1\right)^2 e^{T/\tau_A }\right)}{T \left(f^2 \tau_A ^2+1\right)^2 \left(f^2 \tau_R ^2+1\right)^2 (\tau_A -\tau_R ) (\tau_A +\tau_R )}    \sin (f T)
   \label{eq:S_osc_appendix}       
    \end{align}
\end{widetext}

Taking the limit $T\to\infty$, the above two finite-time contributions in the PSD vanish in the following way,
\begin{equation}
\begin{aligned}
  & \lim_{T\to\infty} \mu_{\mathrm{nos}}(f,T) = 0,
\\
  & \lim_{T\to\infty} \mu_{\mathrm{os}}(f,T) = 0
\end{aligned}
\end{equation}
and $\mu(f,T)$ becomes the stationary PSD $\mu(f,\infty)$.
Expanding Eqs.~\eqref{eq:Delta_S_nonosc_appendix}~and~\eqref{eq:S_osc_appendix} for $f\tau_R\gg1$ and $f\tau_A\gg1$ yields the asymptotic forms presented in Eqs.~\eqref{eq:nonosc_scaling_main} and \eqref{eq:Sosc_main}.

\bibliography{ref}

\end{document}